# Conventional Accelerator Magnets

*S. Sanfilippo*
Paul Scherrer Institute, 5232 Villigen, Switzerland

**Abstract**

This course introduces conventional magnets used in particle accelerators, focusing on both normal-conducting copper coil magnets and permanent magnets (PMs). It covers magnet classification, design principles, material selection, and mechanical constraints. Advantages and limitations of PMs compared to copper coil magnets are discussed. Key construction steps and cooling methods are presented. The course also includes magnetic field measurement techniques and quality control. Practical examples from PSI and CERN illustrate the concepts.

## 1    Introduction

This course provides a comprehensive overview of conventional magnets used in particle accelerators, focusing on their physical principles, design strategies, manufacturing processes, and magnetic field measurement techniques. It begins with a general classification of accelerator magnets—dipoles, quadrupoles, sextupoles, and solenoids—highlighting their respective roles in beam guidance, focusing, and correction. Conventional magnets are defined here as those based either on normal-conducting copper coils or on permanent magnets (PMs). The course distinguishes between these two approaches, presenting the operational principles, integration requirements, and performance characteristics of each type of magnet. Special emphasis is placed on comparing the advantages and limitations of permanent magnet systems—such as energy efficiency and compactness—against conventional electromagnets, which offer tunability and operational flexibility. Material selection plays a central role, particularly in terms of magnetic steel properties. The mechanical design is discussed in relation to constraints such as thermal expansion, Lorentz forces, and alignment precision. The presentation outlines the main steps in magnet development, including the definition of field quality requirements, yoke geometry optimization, conductor layout, and coil fabrication techniques. It also covers practical aspects of magnet assemblies, cooling systems, and compatibility with vacuum environments. The course explores the technological challenges of constructing precise and reliable magnets, supported by examples and case studies from major facilities like PSI and CERN. A final section is devoted to magnetic field measurement techniques—including Hall probes, rotating coils, and stretched wire systems—and quality assurance procedures critical for high-performance magnet deployment.

## 2    Accelerator magnets: classification and properties

### 2.1    Types of accelerator magnets

Magnets are essential components in particle accelerators, enabling the control and manipulation of charged particle beams such as electrons, protons, or ions. Their function is based on the Lorentz force law, where a charged particle (of charge e) moving in a magnetic field (B) experiences a force F perpendicular to both its velocity and the magnetic field:

$$\vec{F} = e\vec{v} \times \vec{B} \tag{1}$$

Different magnets are used for different purposes in particle accelerators, each playing a specific role in controlling and manipulating charged particle beams. Dipole magnets bend particle trajectories, allowing them to follow curved paths within synchrotrons and beamlines. Quadrupole magnets are responsible for focusing or defocusing the beam, particularly in arc sections and at collision points, to improve beam quality and experimental efficiency. Higher-order multipoles—such as sextupoles, octupoles, and decapoles—are employed for beam correction, helping to compensate for chromatic and geometric aberrations. For beam compression or energy selection, systems like dipole-based chicanes are used to sort particles by energy. Solenoids guide and focus beams, especially in transport lines or in experiments requiring precise control of beam shape and direction. Finally, undulators, often made with permanent magnets, are used to generate photons from electron beam oscillations for specific applications like material science and to generate synchrotron radiation by inducing oscillations in the beam. Figure 2 shows pictures of various types of accelerator magnets.

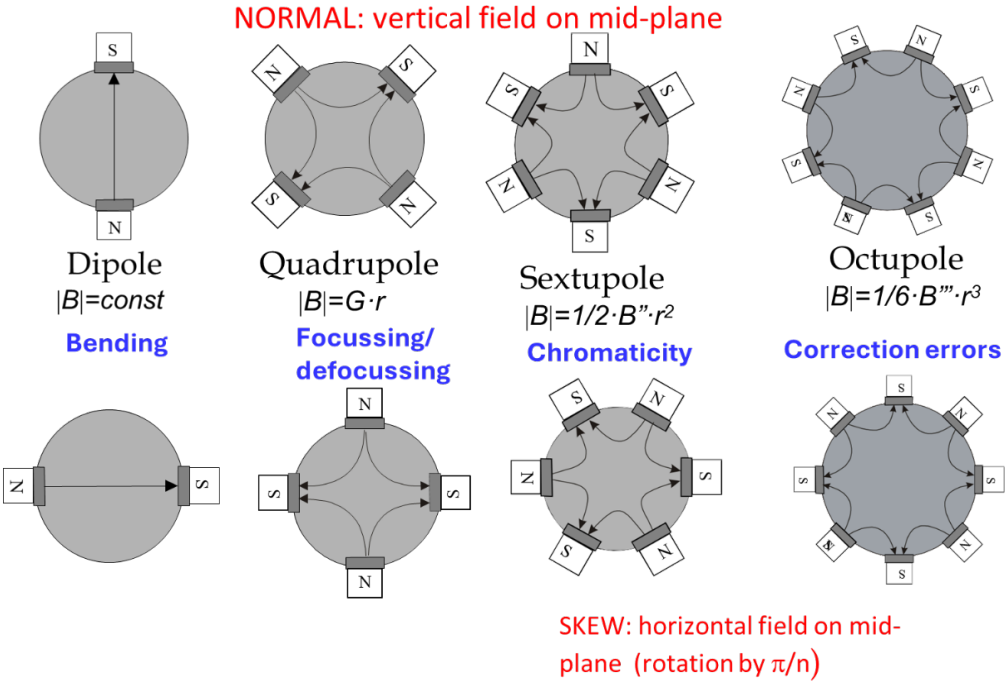

**Fig. 1:** Classification of the accelerator magnets as a function of the pole number- from [4].

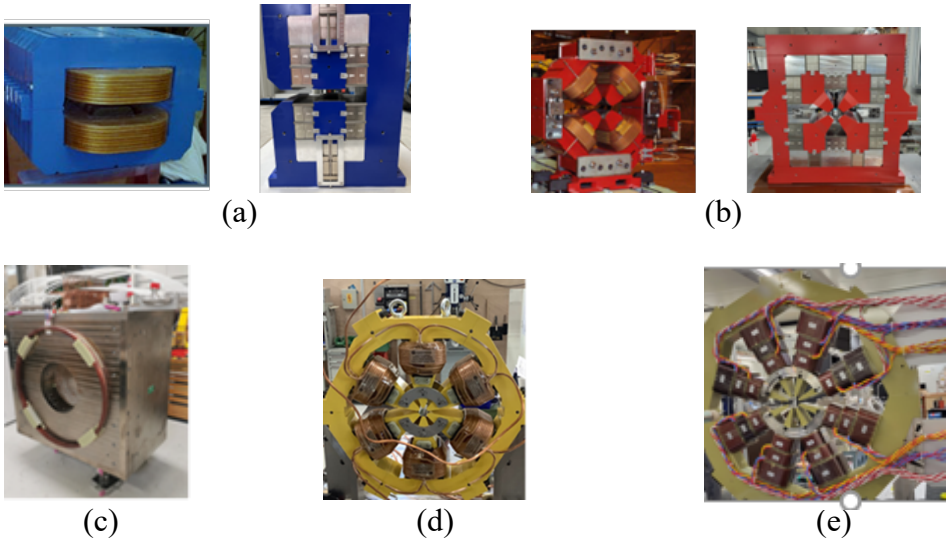

**Fig. 2:** Pictures of (a) Electro and permanent magnet dipole, (b) Electro and permanent magnet quadrupole, (c) solenoid, (d) electro sextupole,(e) electro octupole.

Each magnet type generates a characteristic magnetic field configuration, critical for precise beam dynamics in modern accelerator facilities. In the European nomenclature used for accelerator magnets, the number of magnetic poles is denoted by n. For instance, a magnet with n = 2 is called a dipole, generating a uniform magnetic field from the North to the Sud pole and used for beam bending. A magnet with n = 4 is a quadrupole, producing a field that increases linearly with distance from the center and is used to focus or defocus particle beams. Similarly, n = 6 corresponds to a sextupole, n = 8 to an octupole, and so on. The term "normal" refers to magnets whose poles are arranged to produce a vertical magnetic field in the mid-plane of the beam (typically the horizontal plane in an accelerator). In contrast, "skew" components correspond to the same multipole order, but with the poles rotated by an angle of $\pi/n$. This rotation results in a horizontal field on the mid-plane, introducing a coupling between horizontal and vertical motion, often used for correction purposes. This systematic naming convention helps define both the geometry and the function of the magnet in the accelerator lattice.

The magnetic field distribution in accelerator magnets depends on their type and geometry. In a dipole magnet, the field is constant across the aperture, providing uniform bending to the particle beam; this uniform field is ideal for steering charged particles along curved trajectories. In contrast, quadrupole magnets produce a magnetic field that increases linearly with the distance from the center, focusing or defocusing the beam depending on the orientation. Sextupole magnets generate a field that varies quadratically with the distance from the center, which is useful for correcting chromatic aberrations in the beam optics. Moving further to higher-order magnets like octupoles, the field varies cubically with distance, allowing correction of more complex non-linear beam dynamics. In a solenoid, the magnetic field is primarily oriented along the longitudinal (axial) direction, i.e., along the central axis of the solenoid. The field is approximately uniform and constant inside the solenoid near the center, while it gradually decreases near the ends and outside the coil. Figure 3 shows the space distribution of the field lines different magnet types.

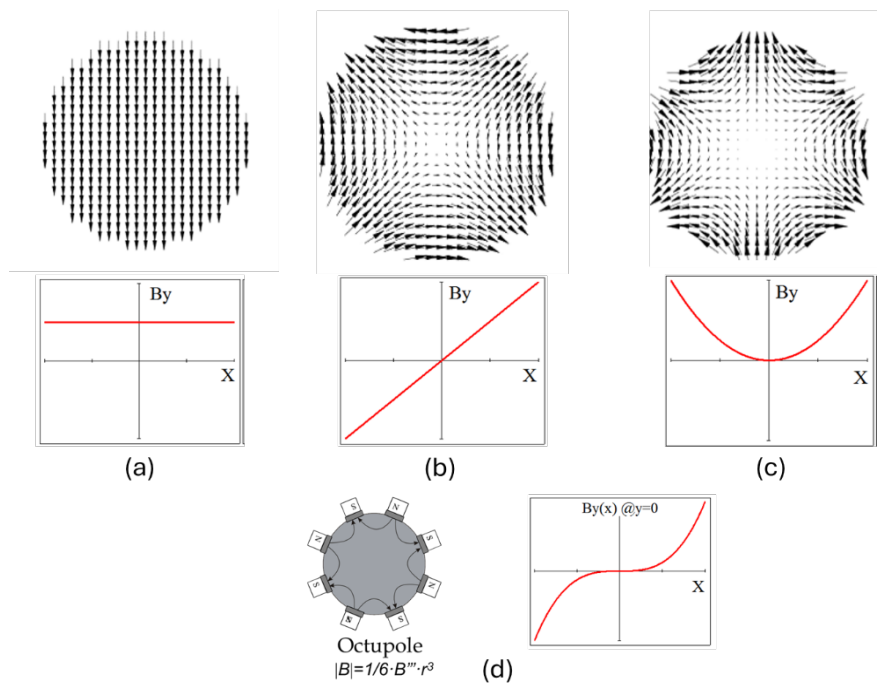

**Fig. 3:** Distribution of the field lines for (a) dipole, (b) quadrupole, (c) sextupole and (d) octupole.

## 2.2 Requirement for accelerator magnets in recent linear colliders storage rings and experimental beam lines

To meet the demanding performance required in modern accelerators, magnets must be designed with a strong emphasis on compactness, sustainability, and low energy consumption, without compromising precision or reliability. These goals are achieved by carefully balancing a wide range of design criteria. Magnets must function reliably under different operation modes—whether continuous, pulsed, or rapidly ramped—and adhere to strict physical constraints, such as limited space, transportability, and weight. The physical aperture must be large enough for the beam to pass, while the field strength (or field integral) must deliver the required deflection or focusing force. A well-defined good field region, typically two-thirds of the aperture radius, is essential to maintain field uniformity and beam quality. Consistent field quality under all operational conditions and accurate alignment of magnetic and mechanical axes are also critical. Furthermore, magnets must be equipped with robust power supplies, effective cooling systems, and be capable of withstanding radiation exposure in harsh environments. Finally, ensuring long-term reliability with minimal maintenance is crucial for sustainable accelerator operations. Figure 4 highlights the various requirements.

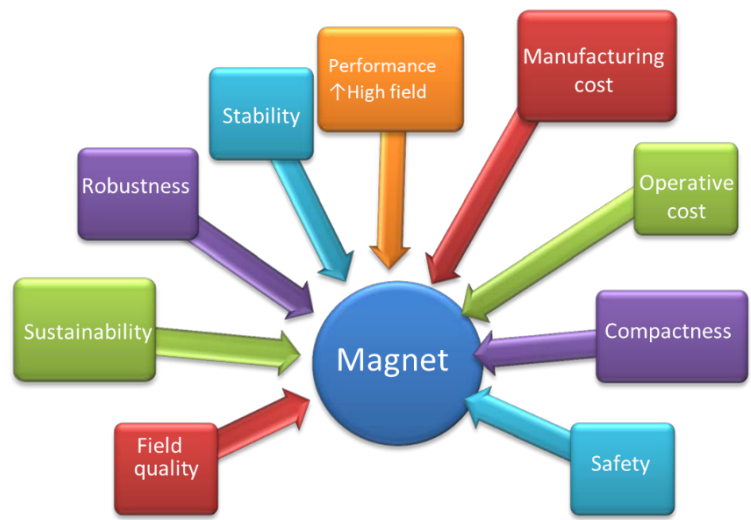

**Fig. 4:** Requirements for accelerator magnets in linear colliders, storage rings and experimental beam lines.

## 2.3 Technology choice

The choice of magnet technology in accelerators depends primarily on the required magnetic field strength and operational environment. For fields below 2 T, iron-dominated magnets are typically used, where the field shape and quality are largely defined by the iron yoke. These can be either normal conducting with copper coils or superconducting, the latter known as super-ferric magnets, where superconducting coils are used but the magnetic field remains shaped by the iron. For higher fields above 2 T, coil-dominated superconducting magnets are required, where the field is mainly generated and shaped by the current in the coils. Permanent magnets, which offer constant magnetic fields without power consumption, are used in low-field applications (typically below 1.4 T) and are best suited for environments with low radiation exposure, as they can degrade over time under high radiation. In contrast, electromagnets—both resistive and superconducting—offer tunability and are therefore widely used in dynamic accelerator applications. A detailed discussion on superconducting magnets will be addressed in a separate dedicated lecture. Figure 5 shows a schematic classification of magnet technologies based on the desired magnetic field strength and operational principle:

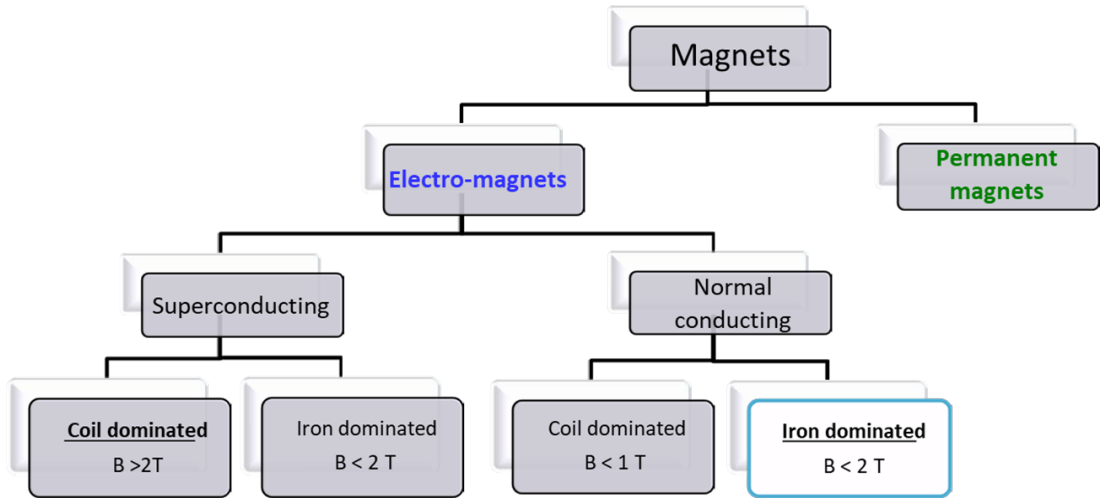

**Fig. 5:** Schematic classification of magnet technologies based on desired magnetic field strength and operational principle. The chart distinguishes between permanent magnets—used for constant fields typically below 1.4 T in low-radiation environments—and electromagnets, which can be either normal-conducting or superconducting. Normal-conducting magnets are further divided into iron-dominated (B < 2 T) and coil-dominated (B < 1 T) configurations [6].

To conclude this chapter, the figures of merit for the main magnet technologies are summarized in Table 1. Each technology—permanent magnets, normal-conducting electromagnets, and superconducting magnets—offers distinct advantages and limitations that must be considered in the context of accelerator design and operation.

Permanent magnet technology is characterized by its simplicity, compactness, and in most cases a lightweight construction. With no need for power supplies, cooling systems, or complex control electronics, these magnets offer extremely low operational costs and minimal maintenance requirements. Their high reliability and robustness make them particularly attractive for applications where long-term stability and minimal intervention are required. These features are especially beneficial in low-field applications—typically in the range of 1 to 1.4 T—and in radiation-free or low-radiation environments. However, their major limitation lies in the fact that they provide only fixed magnetic fields, with limited manual tunability (typically 1–2%) and restricted field strength. In addition, thermal sensitivity and potential demagnetization in radiation environments can impact long-term performance. Normal-conducting electromagnets, usually made with copper coils and iron yokes, provide a flexible and proven solution for many accelerator applications. They allow adjustable magnetic fields—up to approximately 2 T—which can be dynamically controlled to suit varying operational requirements, including beam injection, ramping, and extraction. These magnets are based on well-established, industrial-scale technologies that do not require cryogenics or vacuum insulation. On the downside, they are generally bulkier than permanent magnets, involve higher operating costs due to continuous electrical and water-cooling power consumption, and require regular maintenance to ensure long-term stability. Superconducting magnets stand out for their ability to produce high magnetic fields—often well above 2 T—with excellent field quality and low power consumption once in steady-state operation. They are ideal for compact accelerator layouts or when high beam rigidity is needed, such as in high-energy colliders and very promising candidates for compact gantry systems for medical applications. Furthermore, superconducting magnets can operate in persistent mode, offering highly stable fields over long periods. However, they require sophisticated cryogenic systems, quench protection mechanisms, and careful mechanical and electromagnetic design. These factors contribute to higher initial investment, increased operational complexity, and demanding maintenance protocols. Ultimately, the selection of a magnet technology is not solely driven by performance metrics but must also consider factors such as sustainability, energy consumption, reliability, installation constraints, and long-term operational goals. The optimal choice depends on the specific requirements of the accelerator facility, including field strength, tunability, environmental conditions, and available infrastructure.

Table 1: Pro and cons for the permanent, normal conducting and superconducting technologies in accelerator magnets.

| Type | Advantages | Disadvantages |
|---|---|---|
| Permanent technology | + Compact (no coils, no water pipes)<br>+ No power supply<br>+ Low operation costs (zero power consumption, no water)<br>+ No maintenance (water pipes)<br>+ Less control systems, cables, noise<br>+ High reliability and robustness | − Constant field (tuning: 1-2 %)<br>− Limited in field strength (1-2T)<br>− Permanent magnet magnetization variability (~2 %)<br>− Thermal stabilization needed<br>− Radiation effect (performance)?<br>− Magnetic coupling difficult to compute and correct at the 0.1 % level |
| Normal conducting (electro)magnet | + Flexibility – Variable field<br>+ Moderate field (up to 2 T)<br>+ No need for complicated cryogenic or vacuum systems<br>+ Well know technology | − Larger transverse size<br>− Limited in field (up to ~2 T)<br>− Moderate operating costs for power & water<br>− Maintenance required |
| Superconducting technology | + High fields (>>2T) & Variable fields<br>+ Can be made compact<br>+ Low power consumption | − Complex design<br>− High costs for the manufacturing and cryogenic system<br>− Complicated cryogenics, vacuum, quench protection |

## 3  Permanent magnet technology

This chapter starts by introducing the essential components of an accelerator magnet based on permanent magnet (PM) technology, a solution increasingly adopted for its compactness, energy efficiency, and reliability in low-radiation environments. Unlike electromagnets, PM-based systems rely entirely on the intrinsic properties of ferromagnetic materials—typically rare-earth compounds such as samarium-cobalt ($Sm_2Co_{17}$) and neodymium-iron-boron ($Nd_2Fe_{14}B$). These materials are selected for their high remanent magnetization ($B_r$ ~1.05–1.45 T) and strong intrinsic coercivity ($H_{cj}$ up to 3000 kA/m), ensuring both strong and stable magnetic fields over time. The magnetization curve shown Fig. 6 illustrates their capacity to maintain magnetic field strength ($B_r$) under external demagnetizing forces, with $H_{cj}$ marking the field needed to reverse magnetization—a key measure of magnetic stability [10].

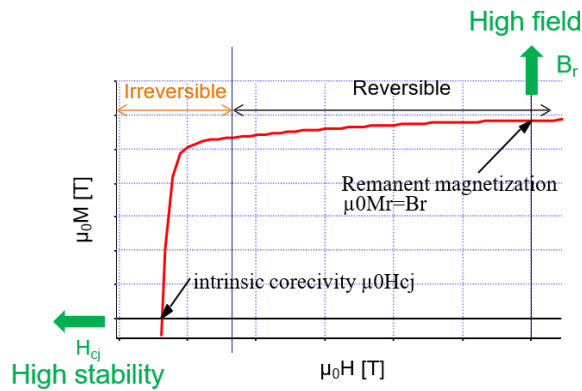

**Fig. 6:** Magnetization variation as a function of the magnetic field for permanent magnet material, showing two important parameters: the magnetic field strength and the intrinsic coercive field, from [10].

The construction of a permanent accelerator magnet (PM) involves several precisely engineered components to ensure stability, field quality, and mechanical robustness. The PM blocks, typically composed of high-performance materials such as $Nd_2Fe_{14}B$ or $Sm_2Co_{17}$, serve as the primary source of the magnetic field. For the 372 SLS2.0 magnets at PSI, typical block sizes are: 30 mm x 47 mm x 54 mm (30 mm: direction of the magnetization). These blocks are embedded within a magnetic circuit that includes a yoke made from low-carbon steel to efficiently guide and contain the magnetic flux. ARMCO-grade high-purity iron is used for the pole pieces to achieve excellent field homogeneity across the beam aperture. To minimize the variation of the magnetization of the blocks with the temperature, several 6 mm nickel-iron (NiFe) strips actings as thermal shunts are integrated at the contact of the blocks. The temperature dependence of the magnetization is reduced by a factor of ten, achieving a ratio of 0.01%/°C. Magnetic shims and moderator plates are also employed to fine-tune the field integral with high precision—typically within a range of 0.01% to 0.05%—so that the integrated field remains as close as possible to the nominal value. Magnetic shims are small ferromagnetic pieces placed near the poles or yokes to locally adjust the magnetic flux, providing fine corrections to field uniformity and the integrated field value. Moderator plates are thin plates that globally influence the magnetic circuit by altering the flux path through the yokes; their position can be adjusted up or down to fine-tune the overall field strength. Aluminum blocks provide mechanical support and ensure precise alignment of all magnetic elements within the structure. Radiation damage on the magnetization is still a subject of investigation, with a reduction of more than 1 % after a cumulated dose of 0.15 MGy. Figure 7 illustrates a practical implementation of this design: a PM-based dipole and reverse-bend quadrupole using $Nd_2Fe_{14}B$ blocks, developed for the triplets in the upgrade of the Swiss Light Source (SLS) at PSI.

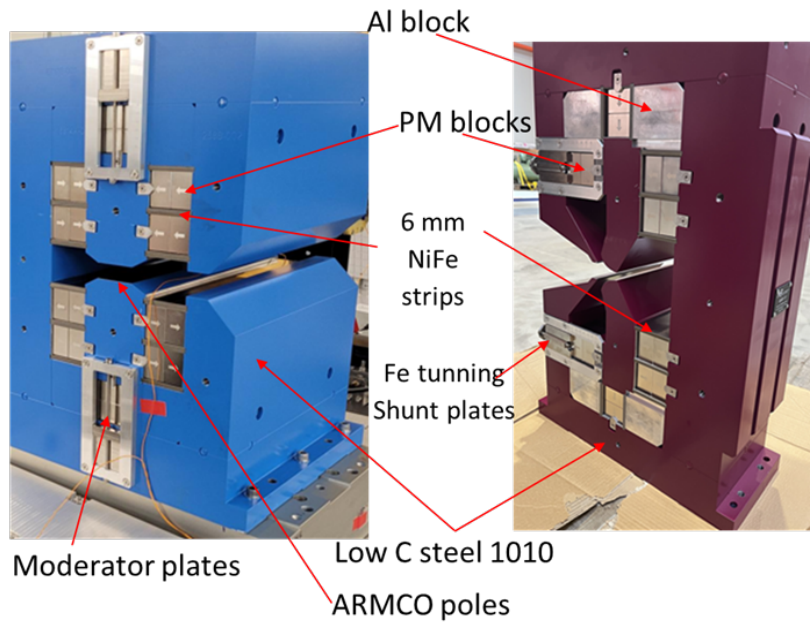

**Fig. 7:** SLS2.0 Dipoles and reverse-bend quadrupoles (quadrupole with a dipole component) made of $Nd_2Fe_{14}B$ with 140 and 96 blocks respectively [13].

For the SLS2.0 upgrade, the management took the strategic decision to assemble all 372 permanent magnet (PM) units in-house at PSI. To ensure safe and efficient assembly despite the strong magnetic pull between blocks (up to 180 kg), a dedicated assembly area was established and equipped with two semi-automatic machines specifically developed for this purpose. These machines allowed precise and secure insertion of PM blocks into the iron yokes, enabling a steady production rate of four magnets per day while guaranteeing mechanical quality and operator safety (Fig. 8).

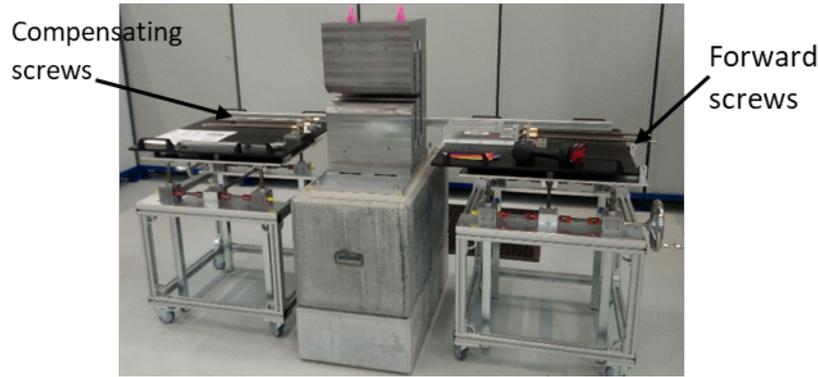

**Fig. 8:** Semi-automatic machine to insert the PM blocks in the iron yoke.

The successful tuning of the field integral of SLS2.0 dipole magnets using precision shims and moderator plates has been achieved. Figure 8 presents the case of 33 dipoles operating at 1.35 T, where initial measurements revealed a standard deviation of approximately 0.2% (35 units), reflecting significant variability due to yoke manufacturing tolerances and fluctuations in the remanent magnetization of the PM blocks (±2%). To mitigate these effects, a dedicated setup using 0.75–1 mm thick shims and adjustable moderator plates was implemented, as shown in the photograph in Fig. 9. The optimization, based on high-precision measurements with the moving wire method, allowed fine adjustment of the magnetic field integral. As a result, the spread in field values was successfully reduced from 0.2% to just 0.01%, demonstrating that the field integral of an SLS2.0 dipole can be tuned with exceptional precision, ensuring excellent reproducibility and magnetic field quality across the full magnet series.

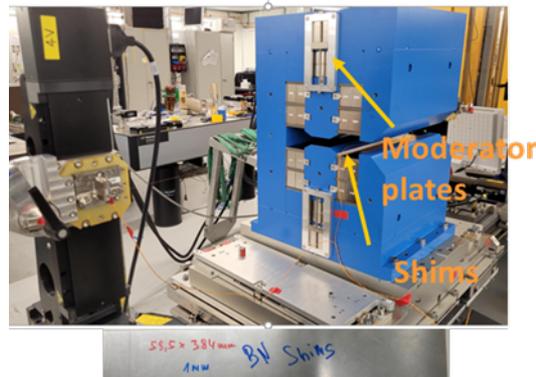

**Fig. 9:** Tuning of the SLS2.0 dipole field integral is achieved using shims for coarse adjustments up to 0.5%, and vertically movable moderator plates for fine tuning with a precision of 0.01%.

## 4 Electromagnets

### 4.1 Overview of the electromagnet production

An electromagnet with copper coils operates by generating a magnetic field through the flow of electric current in the coils. The excitation current, typically carried by insulated copper conductors, produces a magnetic field according to the Biot–Savart law (2). This field is guided and concentrated by a magnetic circuit composed of a low-carbon steel yoke and shaped iron poles, which close the magnetic flux lines (Fig. 9). Due to resistive losses in the copper windings, water cooling is often necessary—especially when the current density exceeds 1 A/mm²—to prevent overheating. The system includes current connections, thermal insulation, and a water-cooling network to ensure reliable and stable operation.

The simplified form of the Biot-Savart law for a circular loop or solenoid is:

$$B = \frac{\mu_0 I}{2\pi R},  \qquad (2)$$

where:

- B is the induction in Tesla,
- $\mu_0$ is the permeability of free space in Henry per meter,
- I is the current through the conductor in Ampere,
- R is the radial distance from the wire in m.

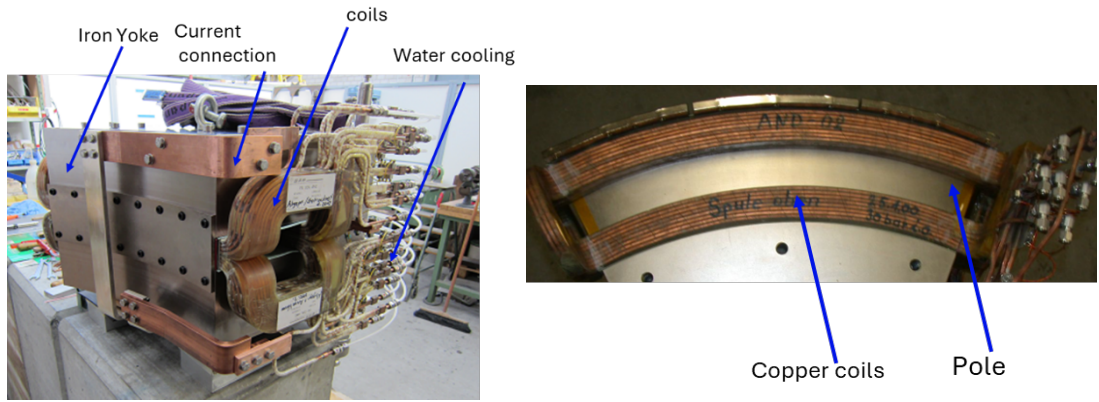

**Fig. 10:** PSI electro-dipole with saddle coil geometry composed of copper coils and low-carbon steel yoke: the right picture shows a sector magnet with tightly wound coils around iron poles, while the left one presents a fully assembled resistive magnet with water cooling, current connections, and insulated copper windings.

The development of accelerator electromagnets is a complex, multi-phase process involving a broad range of technical and operational domains (Fig. 11a). It begins with the input phase, where initial requirements are defined based on the beam dynamics and overall machine layout. This leads into the design and calculation stage, where magnetic, mechanical, thermal, and electrical parameters are computed. These calculations are followed by an engineering and drawing phase, resulting in detailed documentation and CAD models. A prototype including the last mechanical and magnetic design is then fabricated to validate the design through physical testing (electrical integrity, cooling efficiency, field integral and field quality). Once validated, the project moves to series production, where magnets are manufactured in large quantities, often with tight quality control procedures. Following this, the magnets undergo, like for the prototype, power and functional tests to verify electrical continuity, thermal performance, and cooling efficiency. In the magnetic measurement phase, each magnet is precisely characterized using techniques like rotating coils or Hall probes to ensure conformity to field specifications. Once testing is complete, magnets proceed to the installation and commissioning phase, where they are integrated into the accelerator, connected to power and cooling systems, and aligned with beamline components. After successful commissioning, magnets enter the operation phase, where they function continuously under monitored conditions to guide and focus particle beams. During their operational lifetime, magnets must meet rigorous safety, power, and cooling requirements. If the accelerator is upgraded or decommissioned, magnets go through de-installation, followed by storage, disposal, or recycling, depending on their condition and possible future use. Throughout the entire life cycle, a wide array of disciplines is involved. Design, qualification, mechanical assembly, safety, management, and survey all play essential roles. External interfaces such as transport, vacuum, integration, cooling or cryogenics (for superconducting magnets only), and beam optics must be carefully coordinated to ensure full system compatibility (Fig. 11b). The development of accelerator magnets thus requires a systemic approach, with continuous feedback loops between stages to accommodate changes in specifications, operating conditions, or project goals. This interdependent and iterative process ensures that magnets meet the stringent reliability, precision, and performance standards necessary for modern accelerator facilities.

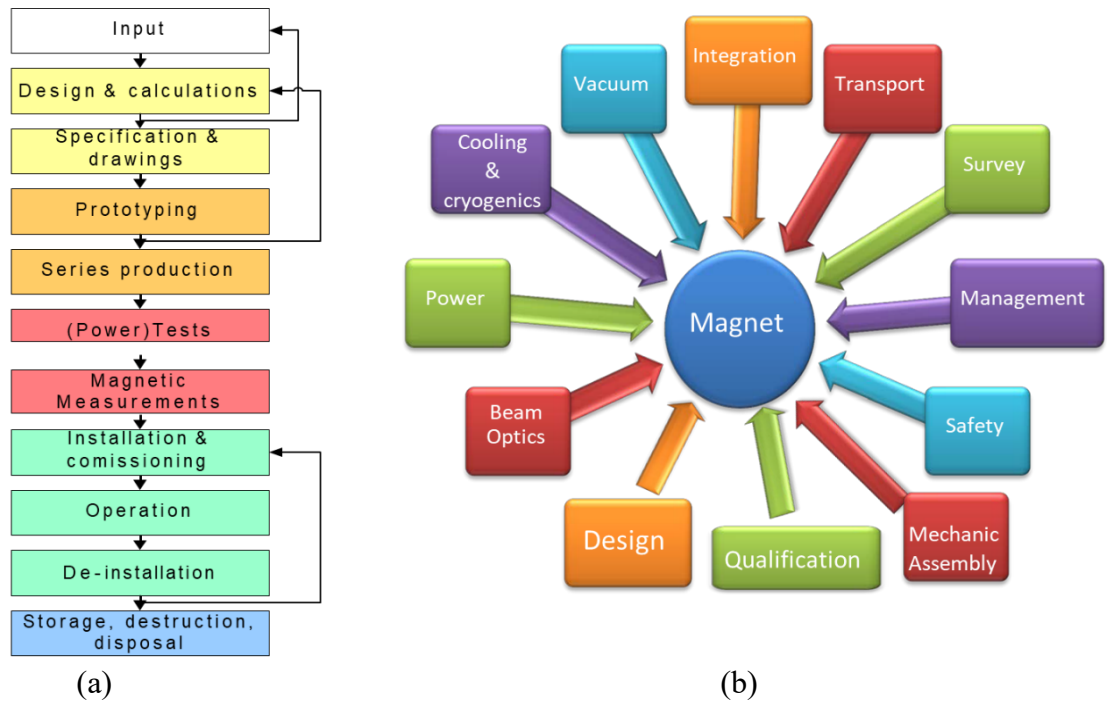

(a)                    (b)

**Fig. 11:** (a): Development process of accelerator electromagnets, from design to disposal.
(b): Magnet life cycle and its integration with key technical domains. Pictures taken from [6].

## 5 Engineering workflow for accelerator magnets: Design, manufacturing, and testing

### 5.1 Overview

The process begins with defining key specifications, which include the following:
- **Physical size:** The overall dimensions of the magnet, including the good field region.
- **Aperture:** The internal opening through which the particle beam will pass.
- **Field integral and homogeneity:** The required magnetic field strength and homogeneity within the good field region.
- **Multipoles:** Acceptable tolerances for magnetic field harmonics to maintain beam quality.

Once the specifications are established, the selection of suitable materials must follow:
- **Conductor:** The material used for the coils, typically copper or aluminum, depending on performance and cost considerations.
- **Insulation:** Electrical insulation materials are chosen to ensure safe and stable magnet operation.
- **Steel:** Magnetic-grade steel is used for the yoke and pole pieces to effectively channel the magnetic flux.
- **Pole and yoke materials:** The choice of geometry and material aims to optimize magnetic performance and mechanical stability.

The next phase involves detailed modeling of the magnet's behavior:
- **Finite Element Model optimization:** Finite element analysis is employed to refine the design, including the coil cross-section geometry, cooling paths, yoke shape, pole geometry, and magnet ends.
- **Thermal and mechanical Modeling:** This step includes optimizing water-cooling paths and verifying mechanical stability under operational conditions.

Coil manufacturing includes several critical steps:
- **Manufacturing Process:** Coils are wound and layered using precise techniques tailored to the magnet design.
- **Insulation application:** Electrical insulation is applied carefully between layers and turns.
- **Impregnation:** The entire coil is resin-impregnated to enhance mechanical strength and electrical insulation.

The manufacturing of the yoke runs in parallel:
- **Tolerance and precision:** Mechanical tolerances are tightly defined to ensure proper field geometry.
- **Machining methods:** High-precision machining is used to shape the yoke and pole components accurately.

Reception tests are conducted after assembly to verify readiness for operation:
- These include visual inspections, as well as hydraulic and electrical testing to confirm assembly quality and integrity.

Magnetic qualification is the final step before installation in the beamline:
- This includes measurements of field strength, field quality, and magnetic axis alignment to ensure the magnet meets all design specifications.

To illustrate the process of an electromagnet production, the case of a dipole used in the bunch compressor of a PSI SwissFEL beamline is presented. The procedure begins with defining magnet specifications derived from beam optics, including parameters such as field strength, aperture, length, and field quality. Numerical simulations are conducted using tools like OPERA 3D™ and COMSOL™ to optimize the magnetic design, followed by mechanical design and prototype fabrication (yoke and coil). Magnetic measurements on prototypes validate the design before finalizing technical drawings and specifications using CATIA. The project proceeds with a call for tender and contract award for manufacturing. If only coils are ordered externally, final magnet assembly is carried out in-house. The completed unit then undergoes reception testing and precise magnetic measurements. The example magnet operates at 200 A with a 0.46 T field, a bending angle of 5°, and a total mass of 200 kg, achieving a field integral uncertainty below 0.1% and field uniformity within 0.05% (Table 2). The pictures in Fig. 12 describes the various steps of the magnet production.

**Table 2:** Dipole specifications for the PSI SwissFEL beam line

| | |
|---|---|
| **maximum current** | 200 A |
| **magnetic flux density** | 0.46 T |
| **bending angle** | 5 ° |
| **yoke dimensions** | L: 250 mm, W: 400 mm, H: 228 mm |
| **Weight** | 200 kg |
| **Uncertainty (field integral)** | <0.1% |
| **DB/B (Good Field Region)** | <0.05 % |

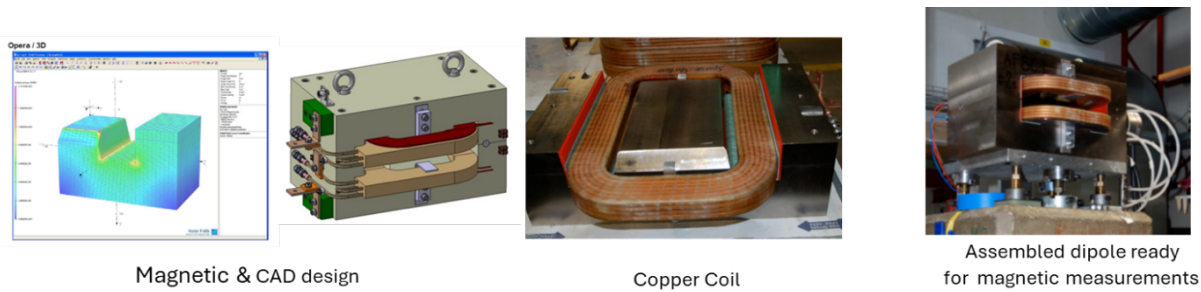

**Fig. 12:** Pictures illustration the various steps for the Swiss FEL dipole design and manufacturing [7].

## 5.2 Magnet design

This chapter aims to describe the main steps of the magnetic design process using finite element analysis (FEA), focusing on its application to the accelerators. While this overview provides the essential concepts and workflow logic, readers are encouraged to consult specialized references for deeper technical implementation [Books and CAS courses in the bibliography].

The first objective in magnet design is to translate the beam optics requirements into a magnetic layout. This is achieved using electromagnetic finite element (FE) models, which enable precise field computation and iterative geometry refinement. These models serve several critical functions: computing the magnetic field distribution and field integral, adjusting the yoke geometry, minimizing regions of high saturation, and tuning pole shapes for desired multipole field components. Beyond field design, FE tools also support coil geometry optimization, material selection, and cost control by minimizing the volume of steel required. Additionally, FE models are used in thermal analysis to estimate heat generation and cooling needs. Several software tools are widely employed in this context. Opera™ (by Dassault Systems) offers commercial 2D and 3D simulation capabilities for a broad range of electromagnetic systems. ROXIE, developed at CERN, is specialized for accelerator magnets and is often used by academic institutions under license agreements. ANSYS™ provides general-purpose 2D and 3D simulations with strong integration for engineering workflows. COMSOL Multiphysics™, known for its user-friendly interface and multiphysics coupling, is also used in electromagnetic modeling and thermal simulations. The selection of a specific tool depends on required accuracy, licensing access, cost, and computational resources.

The finite element method (FEM) relies on discretizing the geometry into small mesh elements—triangles, quadrilaterals, or tetrahedra—on which Maxwell's equations are solved. The accuracy of the simulation depends strongly on mesh quality and solver convergence. Because ferromagnetic materials such as steel exhibit nonlinear B-H behavior, FE solvers must account for magnetic saturation through iterative schemes. The simulation process generally follows three phases: preprocessing, solving, and postprocessing.

**Preprocessing:** The first step in the finite element modeling process is to define the geometry of all relevant regions, including the coils, magnetic yoke (typically made of iron or other ferromagnetic materials), and the surrounding air domain. The coils are then modeled with a specified current density, which serves as the source of the magnetic field in the simulation. To properly channel the magnetic flux, regions made of steel are incorporated into the model, forming the return path for the field lines and influencing the field distribution within the aperture. To optimize computational efficiency and reduce model complexity, symmetry constraints are applied at appropriate boundaries, allowing only a portion of the geometry to be simulated while maintaining accuracy. Finally, nonlinear magnetic properties are assigned to the steel and other materials through the use of B-H curves, which define the relationship between magnetic flux density and magnetic field strength and are essential for accurate modeling of material saturation effects.

**Solving** the electromagnetic field typically begins with the linearized version of Maxwell's equations. Magnetic scalar or vector potentials are computed in the mesh regions. Nonlinear solvers are then used to incorporate the real permeability behavior of steel through successive iterations, refining the solution until convergence is reached. These solvers can be time-consuming due to long meshing and solve times but are essential for high-accuracy models.

**The postprocessing** phase of finite element magnetic simulation includes several essential analyses. First, field line visualizations and graphical outputs are used to display the magnetic field distribution, magnetic flux density, and field strength across different regions of the modeled geometry. These visual tools help to assess the overall behavior of the magnetic circuit and to identify areas of high or low field intensity. Next, the analysis of contours and gradients provides detailed information on field intensity and the degree of uniformity within the modeled region. This is particularly important in applications where a homogeneous field is critical, such as in beam transport magnets or imaging systems. Finally, harmonic analysis, typically performed through a Fourier decomposition of the magnetic field along a circular path, is used to evaluate the field quality. This method identifies the presence and magnitude of undesired multipole components, ensuring that the magnet meets the required specifications for field uniformity and optical performance.

The design process is inherently **iterative**: initial coil and yoke designs are tested, and based on the field results, geometries are adjusted to meet field homogeneity, peak field limits, and field quality targets. This loop continues until a satisfactory design is achieved.

Figures 13a and 13b display two examples of the field distribution of two SLS2.0 magnet: a permanent magnet Triplet (a) and a sextupole magnet calculated using OPERA 3D$^{TM}$.

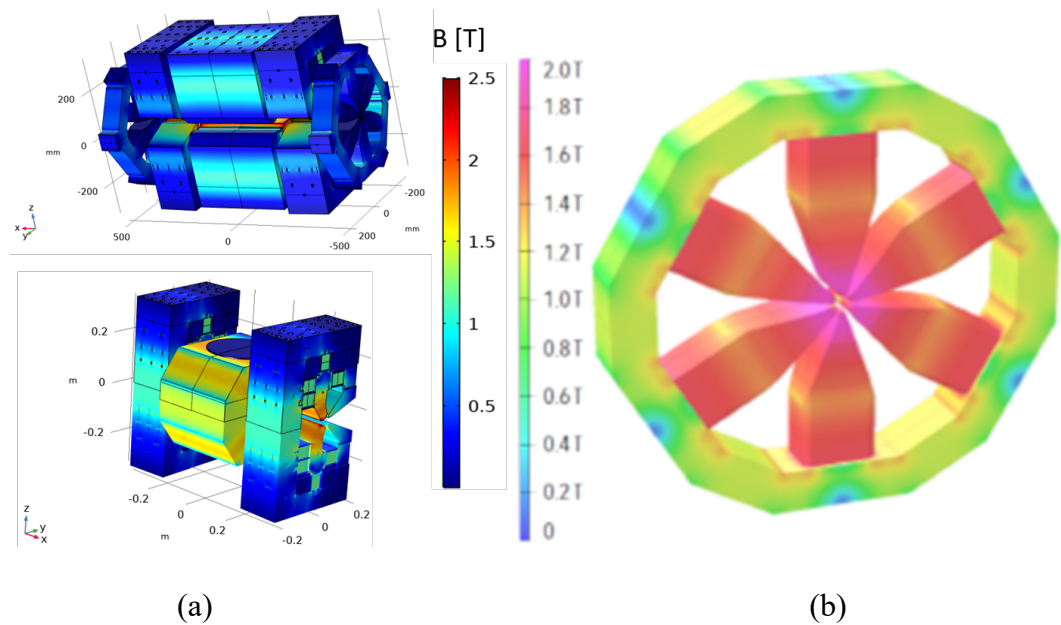

(a)          (b)

**Fig. 13:** Magnetic field distribution of (a) a SLS2.0 permanent magnet triplet (dipole surrounded by two quadrupoles) and (b) a SLS2.0 electro sextupole.

In real-world applications, thermal and mechanical constraints must also be considered. Heat generated by Joule losses in the coils or eddy currents in the yoke must be dissipated effectively, especially for pulsed magnets or superconducting systems. Advanced FE tools allow coupling between electromagnetic and thermal domains to simulate such interactions. In parallel, mechanical deformation and stress due to magnetic forces or temperature gradients can be evaluated in structural solvers, completing the design verification loop.

## 5.3  Water cooled copper coil geometry and manufacturing

The performance, stability, and efficiency of accelerator magnets heavily depend on the design of their coils—particularly the choice of conductor material, the geometry of the coil, and the cooling strategy adopted to manage the significant heat loads generated during operation. In this section, we provide a comprehensive overview of the most commonly used copper-based conductors, their standard geometries, and the technical considerations for implementing efficient water cooling systems.

### *5.3.1  Conductor material: Oxygen-free Copper*

For high-performance electromagnets, the industry standard material for conductors is oxygen-free copper, typically conforming to CA 10200 specifications. This copper grade is selected due to its high electrical conductivity, excellent thermal properties, and enhanced mechanical workability. The absence of oxygen in the copper reduces the formation of oxides, which improves weldability and brazing characteristics—both critical in magnet coil fabrication. Moreover, oxygen-free copper demonstrates greater durability under high thermal and mechanical stress, making it ideal for high-current applications in accelerator magnets.

Key physical properties of recommended used copper include:
- **Purity**: ≥99.95%, ensuring minimal inclusions and high conductivity,
- **Melting point**: 1083 °C, which provides thermal robustness,
- **Resistivity at 20°C**: 1.73 mΩ·cm, indicating very low electrical resistance,
- **Thermal conductivity**: 3.91 W/cm·K, enabling efficient heat dissipation,
- **Density**: 8.94 g/cm³, which impacts both thermal mass and mechanical stability.

### *5.3.2  Coil Geometry: Hollow conductors with cooling channels*

The typical design for accelerator magnet coils involves rectangular hollow copper conductors including an internal cooling water channel (Fig. 14). This integrated channel allows for direct fluid contact with the conductor's inner surface, significantly improving heat transfer. The standard insulation system consists of glass tape wrapping (approximately 0.5 mm thick), often combined with epoxy potting. This not only prevents short circuits but also ensures mechanical integrity and environmental protection.

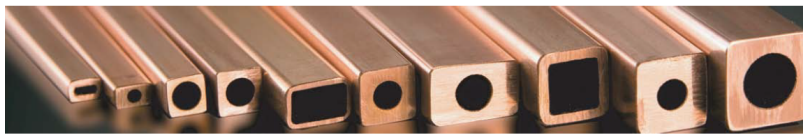

**Fig. 14:** Picture of hollow rectangular copper conductors with various water conduction channels.

Several coil geometries are commonly used (Fig. 15):

- **Racetrack coils**: Simple, elongated loops used in dipoles and some quadrupoles,
- **Saddle coils**: Geometries that allow compact packing and winding stability,
- **Solenoid and tapered coils**: Used in specific configurations for longitudinal field generation or space-constrained systems.

The current density in the copper conductor is defined by the expression:
$$j = NI / A, \qquad (3)$$
where $j$ is the current density, $N$ the number of turns, $I$ the current, and $A$ the total cross-sectional area of the copper.

Different combinations of N and I lead to varying design constraints:

- **Low j** → low power losses and reduced heat generation,
- **High j** → more compact coils and smaller magnets, but higher thermal load,
- **High N** → requires high-voltage, low-current power supplies,
- **Low N** → suitable for low-voltage, high-current systems.

Ultimately, the selection must strike a balance between efficiency, cooling, power supply architecture, and economic constraints. In addition, the number of turns has an influence on the coil size and therefore also on the magnet size.

Figure 15 displays several commonly used geometries:

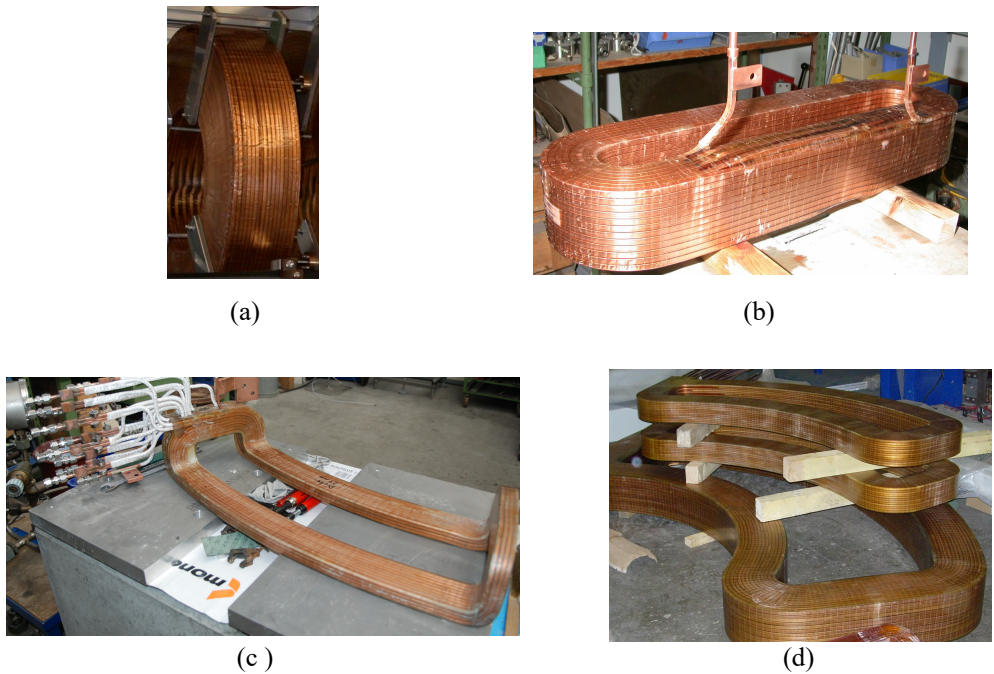

**Fig. 15:** Four common used coil geometries: (a) solenoid, (b) tapered, (c) saddle and (d) racetrack types.

### *5.3.3 Water cooling: Key design considerations*

Heat generation in magnet coils results primarily from Joule losses (FR), and without effective cooling, this heat can damage insulation, deform the copper, or degrade performance. Therefore, most high-current density coils are equipped with water-cooling systems, and the design of these systems follows strict engineering principles.

*5.3.3.1 When is cooling necessary?*

- Coils operating with current densities below 1 A/mm² may dissipate heat passively and may not require forced cooling.
- For coils operating up to 10 A/mm², active water cooling is essential.

*5.3.3.2 Hydraulic design parameters*

- Pressure drop ($\Delta P$): Must be within the range of 10-15 bars depending on the application.

- Flow velocity: Should be above 1–2 m/s to ensure moderate turbulence (Reynolds number > 2000), enhancing convective heat transfer through the boundary layer. However, it must remain below 5 m/s to prevent erosion and vibration-induced fatigue.
- **Temperature rise (ΔT)** across the coil should remain below 30°C to maintain thermal stability.

All these parameters are linked together using the Blasius Law:

$$\Delta P[bar] = 60 L[m] \frac{Q[l\ min]^{1.75}}{d[mm]^{4.75}} \tag{4}$$

*5.3.3.3 Cooling water properties*

- Use of demineralized water: minimal ionic content and thereby reducing the risk of mineral deposits,
- Low resistivity ≥ 0.1 MΩ m, indicative of the water's purity and low ionic concentration,
- Narrow acidic window with pH of 6 to 6.5, optimizing the conditions to prevent corrosion and scaling,
- Filters (reverse osmosis) to remove particles to avoid cooling ducts obstruction.

*5.3.4 Special case of the Mineral Insulated Conductors (MIC) for radiation hard coils*

This chapter is dedicated to the challenges and strategies in designing radiation-hard magnets at PSI particularly for environments where radiation doses can reach up to 10 MGy per day. In such extreme radiation fields, particularly near high-energy targets, the use of conventional materials—especially organic ones like epoxy—is prohibited due to rapid degradation. The pictures of Fig. 16 visually demonstrate the progressive destruction of epoxy under increasing radiation doses, from 5 MGy to 100 MGy, leading to discoloration, structural cracking, and complete disintegration. This justifies the shift towards mineral insulated conductors (MICs), which are entirely inorganic and made of a copper conductor encased in a copper sheath, with magnesium oxide (MgO) powder used as insulation (Fig. 17). MgO cable is radiation-hard but has several critical limitations: it is hygroscopic, fragile, and prone to short-to-ground faults if not handled correctly during fabrication. Moisture absorption is particularly detrimental, as it compromises insulation resistance and increases the risk of electrical failure under high voltage stress.

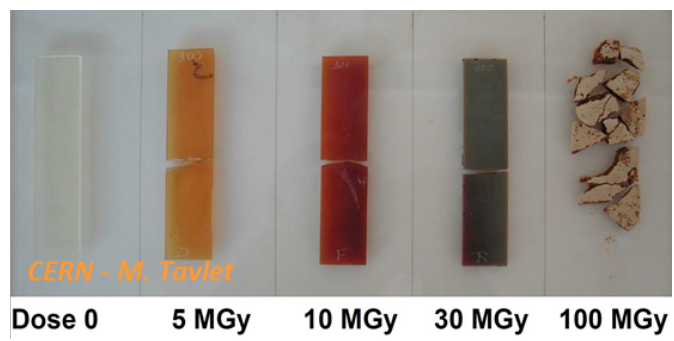

**Fig. 16:** Degradation of copper conductor due to the particle emission from the beam (high energy gamma, neutrons, other hadrons (p+, α, π ...)).

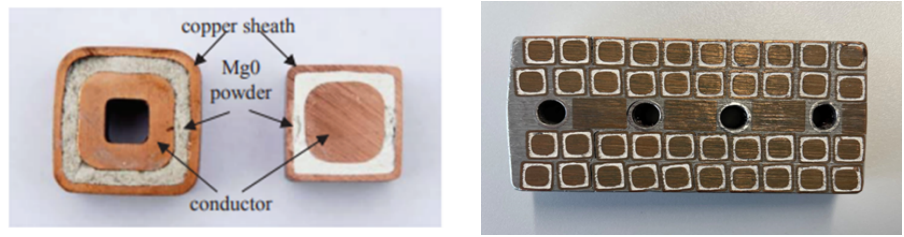

**Fig. 17:** Mineral Insulated conductor (left). Stack of MIC with indirect cooling to avoid corrosion of the copper conductor.

MIC technology requires meticulous care during cable manufacturing and assembly. The MgO powder must remain dry throughout every fabrication step—storage, cable pulling, soldering, and final sealing—since even small traces of humidity can degrade the dielectric strength and promote corrosion in the copper sheath. Another issue is that gamma rays, protons, and neutrons significantly affect copper water-cooling systems. Radiation ionizes water, producing aggressive chemical species (e.g., hydroxyl radicals, hydrogen peroxide) that accelerate copper corrosion (Fig. 18, left). This leads to the formation of copper oxides, which can clog cooling channels and reduce thermal efficiency (Fig. 18, right). Neutrons further damage copper by creating atomic displacements, causing embrittlement and lowering thermal conductivity. Combined with high temperatures, these effects accelerate degradation. To mitigate this, indirect cooling is preferred, isolating copper from direct water contact. The cooling channels are made using stainless steel. The final coil geometry after winding is potted in soft solder to enhance heat transfer between the conductors and the cooling channels and to improve mechanical stability. Voids between the cooling channels can be filled using copper filler pieces to improve heat transfer.

Using radiation-resistant materials like mineral-insulated conductors (MIC) and carefully controlling water chemistry are essential to ensure reliable and long-term magnet performance. Therefore, for commutative radiation doses above 10 MGy, MIC cables are among the only viable insulation solutions, but only if they are perfectly manufactured, hermetically sealed, and indirectly cooled. Indirect cooling minimizes water contact with copper, reducing the risk of copper corrosion.

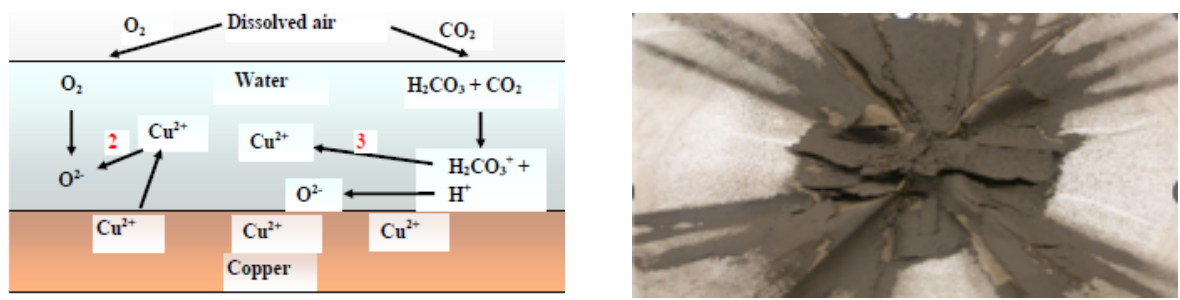

**Fig. 18:** Left - particle beam impact on the acidification of the cooling water. Right - consequence with the obstruction of the cooling channel by the copper oxide.

Ultimately, while MIC provides a path forward for radiation-hardened magnets, its implementation demands a rigorously controlled fabrication environment, vacuum-tight encapsulation, and comprehensive quality assurance testing. Such precautions are vital for maintaining insulation integrity and ensuring the long-term reliability of magnets operating in harsh radiation zones where accumulated doses exceed 10 MGy. A good summary of the challenges of MIC coils and hard radiation magnets can be found in [5].

## 5.4 The importance of the iron yoke

Iron yokes play a central role in the operation and performance of accelerator magnets. Their primary function is to enhance the magnetic field generated by the coils by providing a low-reluctance path for magnetic flux. This effect significantly increases the magnetic field strength in the desired region, particularly in the so-called "good field region." The yoke, typically made of high-permeability materials like low-carbon steel (e.g., 1010 steel), acts as a magnetic circuit element that concentrates and guides the flux. However, the efficiency of this enhancement is limited by the B-H curve and saturation effects. Magnetic steel begins to saturate around 1.5 T, and for B fields below 2 T, the permeability µ can reach values between $10^3$ and $10^4$, allowing the iron to strongly contribute to field enhancement. Above 2 T, the steel saturates, µ approaches 1, and the iron becomes effectively "transparent" to the magnetic field, meaning it no longer enhances it (Fig. 19).

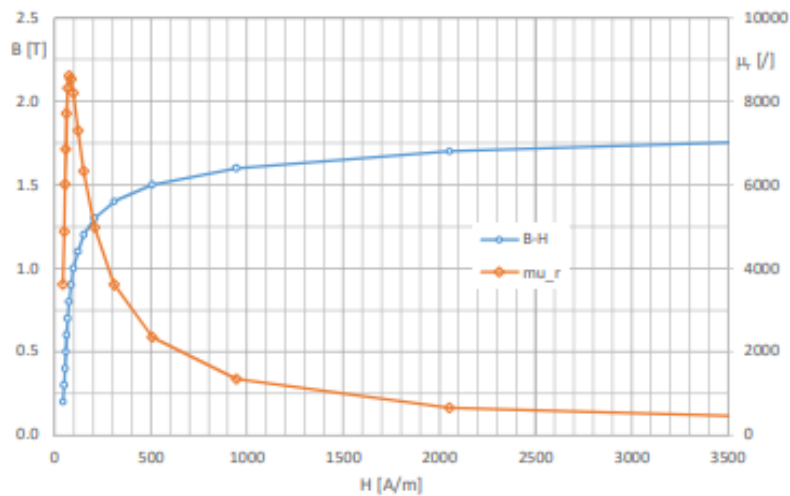

**Fig. 19:** Variation of the magnetic induction B [T] with respect to the applied magnetic field H [A/m)] caused by the iron yoke [12].

In addition to enhancing the magnetic field, the iron yoke plays several other critical roles in accelerator magnets. It contributes to a more uniform distribution of the magnetic field by guiding the flux evenly across the required region, which helps ensure field homogeneity and stable beam dynamics. The yoke also minimizes magnetic leakage by effectively closing the magnetic circuit and containing the field lines within the magnet, thereby reducing stray fields and improving both the efficiency and safety of the system. Furthermore, the iron yoke provides structural support, offering mechanical stability to the magnet assembly by holding the coils in place and maintaining precise alignment under thermal and magnetic stresses. These additional functions make the yoke an essential component in both the electromagnetic and mechanical design of accelerator magnets.

The geometry of the iron yoke is also critical in defining the field quality, mechanical stability, and integration feasibility. Common configurations include C-shaped and H-shaped yokes for dipoles, and several configurations for quadrupoles such as closed, open, or hybrid designs (Fig. 20). C-shaped yokes offer excellent accessibility for mechanical assembly and vacuum chamber installation but suffer from lower field uniformity due to asymmetries in the return path of magnetic flux. H-shaped yokes, on the other hand, provide superior field homogeneity and mechanical rigidity but are more challenging in terms of access. Closed quadrupoles are effective at minimizing magnetic saturation due to the symmetry of the return path, though they require larger coils and are less accessible. Open quadrupole types facilitate easier access for diagnostics and alignment tools, but their complex coil configuration and support mechanics increase manufacturing difficulty.

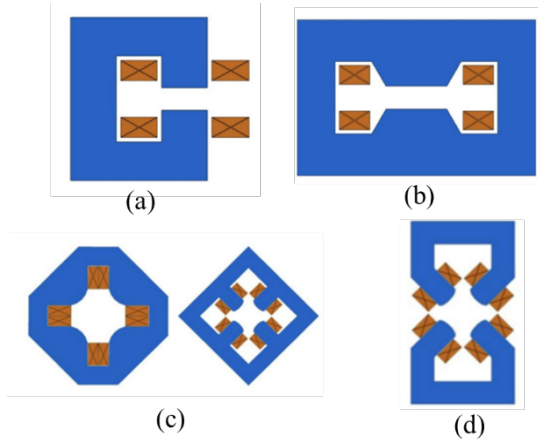

**Fig. 20:** Common shape of iron yoke: (a) C shape, (b) H shape, (c) closed quadrupole, (d) open quadrupole [8]

The choice between solid or laminated iron cores in magnet design depends strongly on the application's operational regime. Solid cores are generally used for steady-state magnets, where the field does not vary over time. These cores offer high magnetic permeability and are easier to manufacture, but they suffer from significant eddy current losses, poor heat dissipation and larger volume, leading to thermal stress and lower field uniformity (~0.1%). In contrast, laminated cores are preferred in time-varying field magnets, such as fast-ramping dipoles or AC quadrupoles. By stacking thin, electrically insulated steel sheets—typically coated with inorganic layers (e.g., phosphating or Carlite) or organic epoxy—laminated cores minimize eddy currents, improve thermal management, and achieve better field uniformity (~0.01%). However, their manufacturing is more complex, and they can be prone to mechanical delamination. In both cases, low-carbon steel (e.g., 1010 steel with <0.1% carbon) is widely used due to its high saturation field and good magnetic permeability. The choice of core type must thus balance electromagnetic performance, thermal behavior, mechanical robustness, and manufacturing constraints, depending on whether the magnet operates in a static or dynamic magnetic environment. Finally, a laminated yoke concept could be more economical for a large series of magnets, since the stamping and assembly process is more efficient than individual precision machining of every single yoke piece.

The selection of yoke material is critical due to its impact on magnetic permeability and saturation characteristics. For low-carbon steel, such as 1010 steel, with carbon content typically below 0.1%. This material ensures high permeability and a relatively high saturation field. However, even with careful material control, variations in magnetic properties can arise due to batch inconsistencies. For example, studies conducted at CERN over 50,000 tons of steel showed significant variation in coercivity ($H_c$) and permeability, even within the same steel batch. As illustrated by measurements on ARMCO steel samples (Fig. 21), the permeability μ can vary by around 1% across samples from the same batch. Across large-scale production, absolute uncertainties of 3–5% in permeability are common. These variations can impact the magnetic field quality and must be accounted for during magnet design and field tuning.

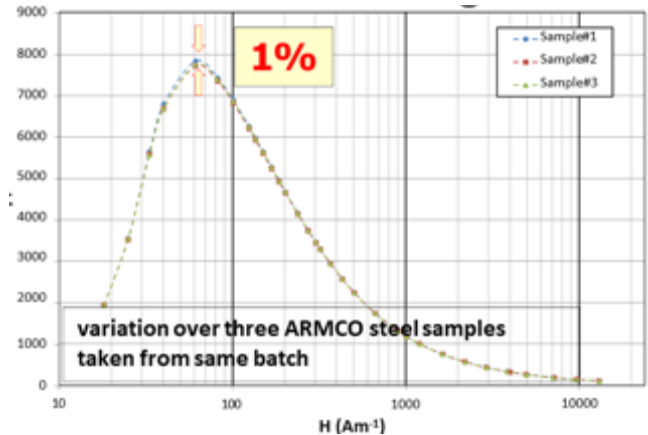

**Fig. 21:** Variation of magnetic permeability within three ARMCO steel samples [9]

# 6  Magnetic qualification

## 6.1  Sources of magnetic field errors

Magnetic field imperfections and instabilities in accelerator magnets arise from a range of factors, both static and dynamic, that must be carefully understood and mitigated during design, manufacturing, and operation. The source of magnetic field errors are numerous:
- Manufacturing errors and mechanical tolerances, typically in the range of 20–50 micrometers for conventional electromagnets and permanent magnet yokes, lead to deviations from the ideal magnetic geometry. In superconducting magnets, coil geometry errors relative to the design configuration can introduce significant asymmetries in the field distribution.
- Material-related uncertainties, such as variations in permeability of yoke steels or remanent magnetization in permanent magnets, typically within a few percent, affect the field quality. Furthermore, the hysteretic behavior of magnetic materials can result in non-reproducible magnetic states, particularly during cycling or polarity changes.
- Finally, magnetic and mechanical deformations due to Lorentz forces and thermal contraction introduce small but significant changes in alignment and coil shape, degrading field stability and reproducibility. Addressing these sources of error requires precise manufacturing, detailed modeling, and dedicated field-correction strategies.
- When magnets are operated in high-density environments or designed for multi-functionality, magnetic coupling between adjacent elements can distort the local field.
- Dynamically, eddy currents generated during fast current ramps can induce transient field errors and energy losses. In superconducting systems, persistent current hysteresis—arising from the non-linear magnetization linked to the critical current density ($J_c$)—can distort the field, especially at low fields.
- Additionally, phenomena such as current redistribution and long-term magnetization decay at cryogenic temperatures (e.g., 1.9 K) further alter the field over time. This error source needs to be considered for superconducting wires and cables.

## 6.2  Objectives

Measuring the magnetic field in the good field region is essential to ensure that accelerator magnets perform according to the strict tolerances required by the machine optics. One of the primary motivations is to carry out acceptance tests, which verify that each magnet respects the design specifications, especially in terms of field uniformity and alignment. These measurements also provide essential fiducialization data that allows for precise installation and alignment of the magnets within the accelerator tunnel.

Furthermore, magnetic measurements are indispensable to accelerate the commissioning of the machine, as they provide a wealth of input data that can replace time-consuming beam-based alignment procedures in the early stages.

Additionally, measurements are used to validate magnetic models derived from finite element simulations and analytical calculations, which cannot fully capture the real-world behavior of the materials and assemblies. This confirmation step is vital, as the simulations typically target an accuracy better than 0.01%, which is extremely difficult to achieve due to several factors. First, geometrical uncertainties, such as mechanical tolerances or assembly misalignments, can easily exceed 50 micrometers, contributing to approximately 0.5% field error. Second, material property uncertainties, particularly in magnetic permeability of yoke steel like ARMCO cores, or remanent fields in permanent magnets like NdFeB , can introduce an additional 1–3% variation (see 5.3). Finally, magnetic coupling between magnets or within multi-function systems can be complex and time-consuming to simulate with sufficient accuracy.

For all these reasons, experimental measurements are necessary not only to determine the magnetic parameters used for beam simulations, but also to monitor the quality and consistency of the magnets during series production and to steer corrections during manufacturing. Without such experimental verification, the actual performance of the accelerator magnets would remain uncertain, risking suboptimal beam control or machine instabilities. Thus, magnetic measurements are not only a diagnostic tool but also a mandatory quality assurance step to ensure reliable and reproducible magnet behavior in demanding accelerator environments.

## 6.3 Field measurement outputs

*Field integrals:* The main goal is to determine the magnetic field integral, such as ∫B·dl for dipoles, ∫g·dl for quadrupoles, or ∫$B_3$·dl for sextupoles, which represents the integrated magnetic strength along the effective magnetic length that the beam experiences, including fringe and stray fields. This value is critical for precise beam trajectory control, momentum selection, and lattice tuning.

*Multipoles:* Another major aim is the evaluation of field quality, which quantifies the deviations of the actual magnetic field from the ideal design field. The multipole decomposition of the magnetic field in accelerator magnets is a mathematical formalism used to describe the magnetic field inside a cylindrical aperture, where the beam travels. In this 2D approximation (valid due to the long length of the magnet relative to its transverse dimensions), the magnetic field components Bx and By (main direction of the field) are expressed as a complex series:

$$B_y + iB_x = \sum_{n=1}^{\infty} C_n \left(\frac{s}{R_{\text{ref}}}\right)^{n-1}, \quad (5)$$

where $C_n = B_n + iA_n$ are complex multipole coefficients.

These coefficients decompose the total field into **normal components** $B_n$ and **skew components** $A_n$, corresponding respectively to cosine-like and sine-like field contributions [1].

$$B_y + iB_x = 10^{-4} B_1 \sum_{n=1}^{\infty} (b_n + ia_n) \left( \frac{x+iy}{R_{ref}} \right)^{n-1}.$$ (6)

These coefficients are commonly normalized by the main field:

$$b_n = \frac{B_n}{B_{ref}} \qquad a_n = \frac{A_n}{B_{ref}}$$

Each term in the series corresponds to a specific multipole order:

- n=1: dipole (normal $b_1$, skew $a_1$),
- n=2: quadrupole (normal $b_2$, skew $a_2$),
- n=3: sextupole (normal $b_3$, skew $a_3$),
- and so on...

These harmonics describe deviations from the ideal field and are fundamental to beam dynamics since they introduce field imperfections (nonlinearities, coupling) that affect the beam quality. These coefficients are often scaled by $10^4$ to express the deviation from the ideal field as a fraction of 0.01% called a unit (1 unit =0.01 % of the normalized field error). The normalized multipoles $b_n$ and $a_n$ are a-dimensional and provide a standard way to compare field errors across different magnets or configurations.

$$b_n(unit) = \frac{B_n}{B_{ref}} * 10^4$$

Thus, multipole decomposition is essential for field characterization, magnet acceptance, and beam optics design, as it allows a precise quantification of field quality and harmonic distortions in accelerator magnets. The chapter 4 of the book of J. Tanabe (ref. in the bibliography) provides a comprehensive overview of the links between mechanical imperfections and the presence of various multipoles.

*Field mapping:* Field maps in 2D or 3D, using point-like or continuous line scans, are employed to assess spatial field homogeneity across the aperture and to identify gradients or asymmetries that may affect beam stability.

*Magnetic axis:* Determining the magnetic axis location in multipole magnets (n ≥ 2) is crucial, where the locus of the zero-field point defines the reference trajectory. This axis is measured relative to external mechanical or optical references to ensure precise alignment within the accelerator, often through fiducialization. These measurements allow for accurate installation, alignment, and correction of magnets within the beamline. Ultimately, the goal is to achieve a high degree of reproducibility and predictability in the magnetic environment, thereby enabling optimal beam dynamics, minimizing emittance growth, and ensuring a stable operation of the accelerator.

## 6.4 Field measurement techniques

To achieve these objectives, combined techniques—cross-checking field integrals, maps, quality, and axis location—are used with rigorous calibration and alignment procedures. Selecting an appropriate magnetic field measurement technique is a complex task that depends on several critical factors related to the application, environment, and desired precision. First, the type of quantity to be measured—whether it's a specific field component ($B_x$, $B_y$, $B_z$), the field integral, or a full field map—determines the most suitable sensor and acquisition system. Second, field characteristics, such as the expected strength (from µT to several Tesla), whether the field is DC or AC, and the required uniformity, play a major role in narrowing down the applicable technologies. For instance, Hall probes, fluxgates, NMR sensors, and SQUIDs all operate over different field ranges and accuracies. The accuracy requirement—ranging from 1% to as low as 0.001% (10 ppm)—also drives the choice, with high-precision applications necessitating more sensitive and often more expensive setups. Additionally, physical access to the region being measured must be considered, as it influences the size and type of probe, as well as the ability to position it reproducibly inside the magnet aperture. Environmental conditions like temperature (e.g., room temperature or cryogenic operation) and magnetic noise may affect sensor behavior and require specific compensation techniques. Finally, practical constraints such as measurement time, cost, and available manpower influence the feasibility of the selected method. The graph in Fig. 22 displays the domains of use of several measurement techniques following the required uncertainty and the magnetic field intensity [11].

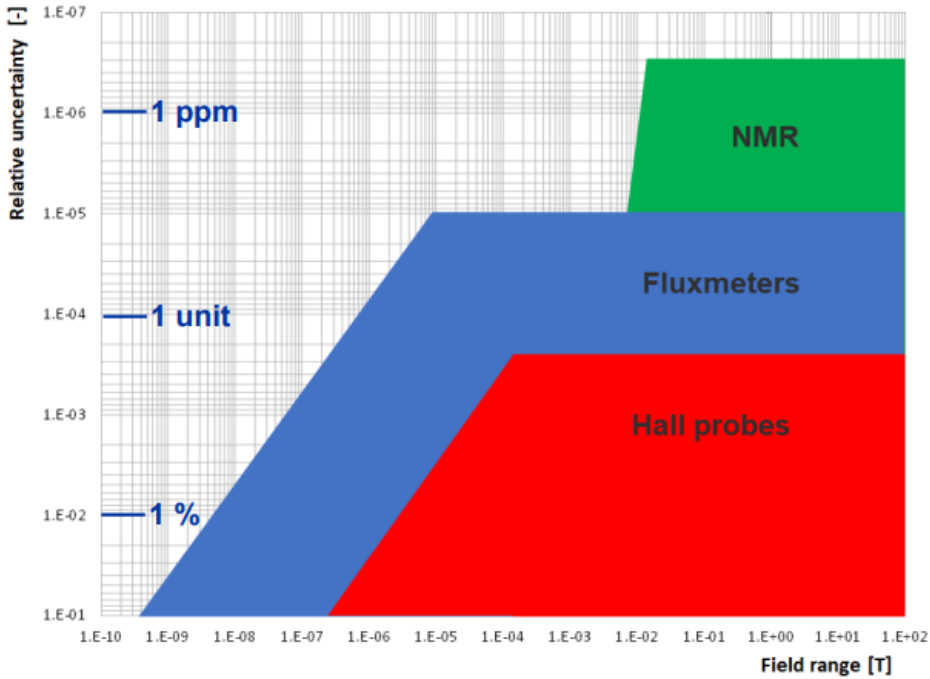

**Fig. 22:** Uncertainty vs field range for several measurement techniques [11]

The diagram of Fig. 23 illustrates the simplified measurement chain used in magnetic field characterization, emphasizing that the primary measured quantity is a voltage, which must be carefully converted into magnetic field information.

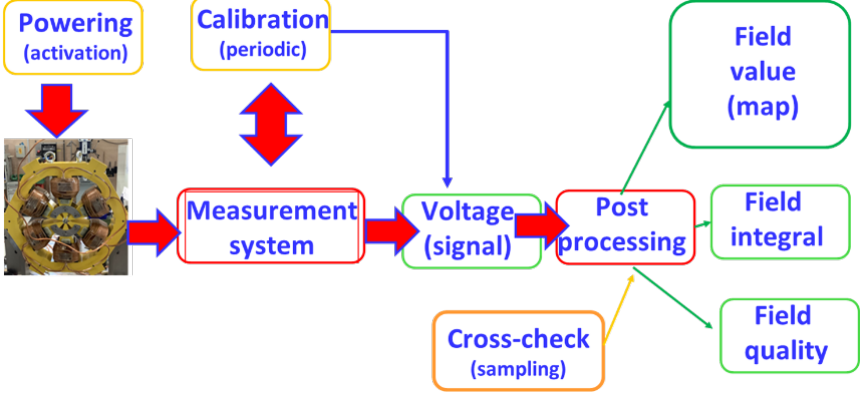

**Fig. 23:** Magnetic measurement chain (case of an electromagnet) from powering to the resulting magnetic measurement outputs.

The process begins with powering the magnet, generating the field to be measured. The signal is captured by a measurement system (such as a Hall probe, rotating coil, or fluxgate), which puts out a voltage proportional to the magnetic field. This voltage is not directly the magnetic field—it must be converted through calibrated relationships that depend on sensor characteristics (e.g., geometric factor, Hall coefficient, gain, and orientation). Calibration is essential and must be performed regularly to ensure accuracy and to compensate for sensor drift or temperature effects. The resulting voltage signals then undergo post-processing, where software or signal processing techniques extract meaningful quantities such as the field value, the field map, the field integral, or the field quality in terms of multipoles or uniformity. Additionally, cross-checks using reference magnets, sampling devices, or complementary methods are often implemented to validate the results and assess measurement reproducibility. This entire chain—from activation and sensing to signal interpretation—ensures that magnetic field measurements are accurate, traceable, and reliable.

**Field mappers** represent the first and most widely used family of magnetic measurement systems for characterizing accelerator magnets. These systems are designed to map the spatial distribution of the magnetic field inside the aperture, providing precise measurements of the magnetic field vector (typically $B_x$, $B_y$, $B_z$) across a defined grid or along specific paths. The core of a field mapper is usually based on Hall probes, which convert the local magnetic field into a voltage signal via the Hall effect. As shown in the diagram, a DC current I flows through a thin semiconductor slab (e.g., InAs, InSb, GaAs), and under the influence of the Lorentz force from the magnetic field B, a transverse Hall voltage $V_H$ is generated (Fig. 24). This voltage is proportional to the magnetic field component perpendicular to the current and is related through the expression:

$$V_H = G \cdot R_H \cdot I \cdot B \cdot \cos(\theta), \qquad (7)$$

where G is the geometric factor and $R_H$ is the Hall coefficient. To ensure accuracy, the output voltage must be converted into magnetic field strength using a carefully calibrated relationship, accounting for sensor-specific parameters and alignment. Field mappers can be one-dimensional (scanning along a line), two-dimensional (mapping a plane), or three-dimensional (scanning the full volume), and are essential for determining field homogeneity, magnetic axis position, and effective magnetic length. Thanks to their versatility, reliability, and compatibility with post-processing techniques, field mappers based on Hall sensors remain a cornerstone in magnet acceptance testing and beam optics tuning in modern accelerator facilities. A comprehensive description of the multiple types of Hall sensors and applications to magnetic measurements can be found in [3].

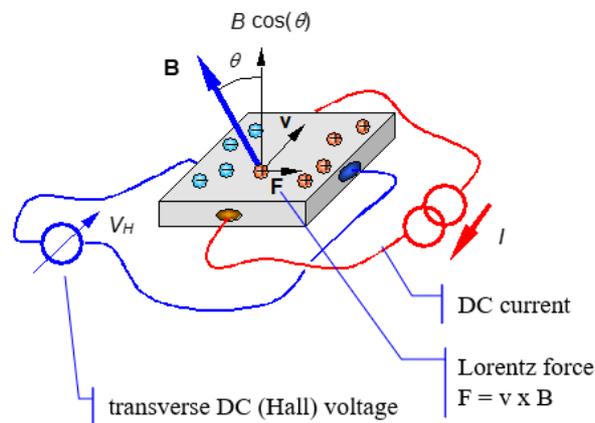

**Fig. 24:** Principle of Hall Effect in a Semiconductor Sensor

**Fluxmeters** [2] represent a second key family of magnetic field measurement systems, are commonly used in accelerator magnets and complementary to Hall probe-based field mappers. These systems rely on **Faraday's law of induction**, which relates the induced voltage $V_C$ to the **rate of change of magnetic flux** through a surface A.

$$V_C = -\frac{d\Phi}{dt} = -\frac{d}{dt}\iint_A \mathbf{B}\cdot\mathbf{n}\,dA = -\iint_A \frac{\partial \mathbf{B}}{\partial t}\cdot\mathbf{n}\,dA - \oint_{\partial A} \mathbf{v}\times\mathbf{B}\,d\ell$$

- Faraday's law (total derivative)
- fixed-coil, time-varying field
- coil rotating, translating (wire)

(8)

The induced voltage is captured with a fixed coil in a time varying field and/or by a rotating coil or a translating wire for DC field. This principle enables the measurement of magnetic fields and their integrals through voltage integration, allowing one to extract both the main field component and field harmonics via Fourier analysis. Unlike Hall probes, which give a local point-like field measurement, fluxmeters offer global information over the coil area or trajectory.

There are two types of fluxmeter systems: pickup coils and wire loops.
- Pickup coils can either rotate within a static magnetic field (changing the surface orientation and sweeping through the flux), or remain fixed in place while the magnetic field varies in time (e.g., during magnet ramping). In both cases, the variation in flux induces a voltage signal proportional to $\partial B/\partial t$.
- Wire loops, typically a single turn of wire, are moved translationally or rotationally through the magnet aperture. As the wire cuts through magnetic field lines, it samples the flux through its changing surface area. These wires are widely used at facilities like PSI to measure the magnetic axis, integral field, and harmonic content of quadrupoles and dipoles.

The output voltage from these systems is related to the pick-up surface depending on whether the coil is rotating, moving linearly, or seeing a time-varying field.

$$-V_c = \frac{\partial \Phi}{\partial t} = \begin{cases} A_c \dot{B} \\ A_c B \omega \\ A_c \nabla B v \end{cases}$$

$A_c$: pick-up coil surface
$\omega$ rotation speed

Field multipoles can be obtained by measuring the voltage induced in a rotating coil or wire as it moves through the magnetic field of an accelerator magnet. This voltage, generated according to Faraday's law, reflects the time variation of the magnetic flux through the coil as it samples the field across different angular positions. By discretely recording the voltage over a full rotation, a periodic signal is obtained, which can be analyzed using Fourier decomposition. The resulting Fourier components correspond to the main field and multipole content of the magnetic field (see Section 6.3).

Rotating coil systems made of printed circuit board (PCB) coils (e.g., PSI/Elettra/CERN designs, Fig. 25) are widely used for harmonic analysis in quadrupoles and sextupoles. Moving wire systems are efficient for field integral measurements and for aligning magnets by identifying the magnetic center with high precision. The various magnetic measurement processes are explained in [9].

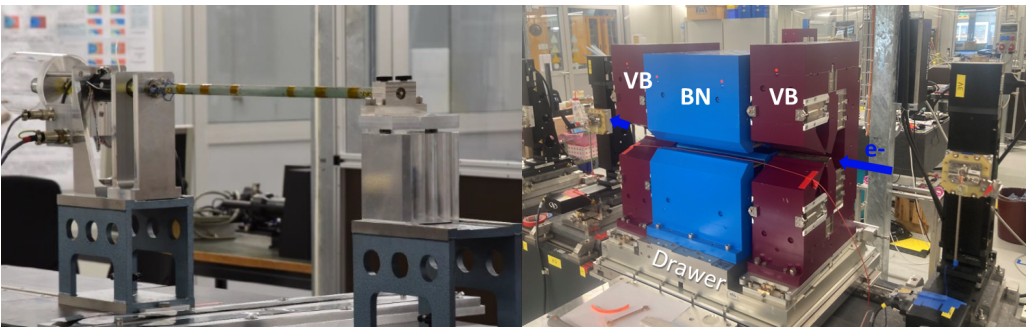

**Fig. 25:** Left - 400 mm long-rotating coils (PSI/Elettra/CERN) made of 5 PCB coils; Right - Field Integral of the SLS2.0 Permanent Magnet triplet (VB = quadrupole, BN = Dipole) using a moving wire.

## 6.5 Field quality tuning in conventional magnets

Tuning the magnetic field quality in conventional magnets is a delicate process that primarily depends on the design and shape of the magnet pole, as well as additional geometric modifications. The designer begins by predicting the magnetic field produced by a specific pole profile using simulation tools and then iteratively adjusts the geometry to achieve the desired field uniformity and harmonic content. A key technique involves the use of shims—small pieces of ferromagnetic material added near the edges of the poles (Fig. 26). These shims compensate for the finite truncation of the pole and help suppress "allowed" harmonic errors, which arise due to the finite extent of the magnetic pole. The shape and placement of these shims influence how the field rises and falls at the pole edge, thereby modifying the amplitude of certain harmonic components. In dipoles, for instance, the allowed multipole errors are of the form n=3,5,7 while in quadrupoles they are n=6,10,14,… due to the underlying magnetic symmetry. The goal of shim tuning is to minimize these components and improve field homogeneity across the aperture.

Another important strategy is the use of chamfers, which are cuts or bevels applied at the ends of the pole to control the longitudinal variation of the magnetic field and better define the effective magnetic length. Chamfers also help preventing saturation of the iron at the pole ends and improving the reproducibility of the field integral. While allowed harmonics can be systematically addressed by design, non-allowed harmonics—including both normal and skew components—often originate from practical imperfections such as mechanical misalignments, assembly errors, or material inhomogeneities. These must be corrected by precision machining, fine pole shimming, or magnet sorting strategies.

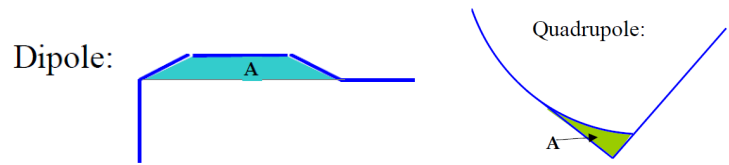

**Fig. 26:** Shims to tune the field quality of dipoles (left) and quadrupole (right)

## 6.6 Magnetic coupling (cross-talk)

Magnetic coupling between neighboring magnets—also referred to as magnetic crosstalk—can significantly impact the field integral and field quality, especially in permanent magnet (PM) systems where post-installation tuning options are limited. This effect arises when the fringe fields or magnetic return paths of adjacent magnets interfere with one another, altering the internal field distribution in each magnet. Magnetic coupling can occur between two permanent magnets, or between a PM and an electromagnet or superconducting magnet, and the resulting disturbances can vary from a few percent up to 20% in extreme cases. To mitigate these field errors, action must be taken at two levels: The first level is during the magnetic design phase, where simulations must go beyond isolated magnets and include full assemblies with realistic neighboring magnets to accurately predict field changes caused by coupling. Electromagnetic simulation tools (like OPERA®) should model not only the target magnet but its environment to capture the mutual influences. The second level involves experimental cross-checks, where selected magnets are measured in situ or in pairs to confirm the predicted effects of magnetic interaction. These measurements help fine-tuning the models and allow for correction strategies, such as mechanical shimming or optimized pole shaping, to be applied upstream. The case of the influence of the SLS2.0 sextupole SXQ on the field strength of the permanent dipole BE is underlined in the following (Fig. 27). The nearby electromagnetic sextupole SXQ reduces the measured field integral of the permanent magnet dipole BE by approximately 4.6%, compared to 3.8% predicted in simulation. The BE permanent dipole series was fine-tuned additionally using shims to achieve the required field integral before their installation in the tunnel.

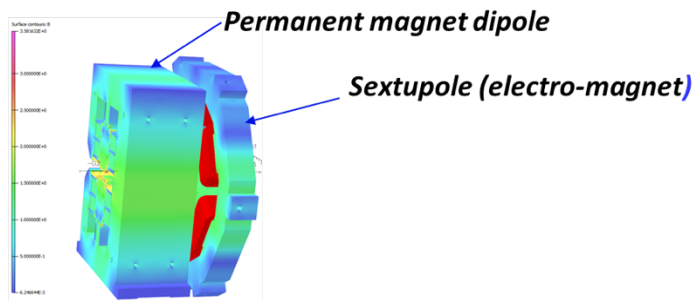
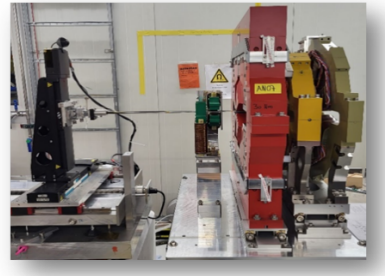

**Fig. 27:** Magnetic design of the permanent magnet dipole, taking into account the neighboring electro sextupole (Left). the dipole field integral was increased by 3.8 % with respect to its nominal value. Experimental check of the field integral reduction by the sextupole based on Hall probe measurements (Right). The measured coupling effects show a reduction of 4.5 %. The series of permanent dipole BE was further tuned with shims before their installation in the tunnel.

# 7 Conclusion

This lecture has provided a short overview of conventional accelerator magnets, focusing on both normal-conducting electromagnets and permanent magnet systems. We explored the fundamental principles behind dipoles, quadrupoles, and higher-order multipoles, along with their essential functions in beam steering, focusing, and correction. The advantages and limitations of copper coil-based magnets were examined in relation to tunability, thermal performance, and operational flexibility, while permanent magnet assemblies were discussed for their compactness and energy efficiency. A particular emphasis was placed on the design process, including magnetic modeling, yoke geometry, coil technology, and field tuning techniques such as pole shaping, shimming, and chamfering. The lecture also included the specific case of Mineral Insulated Conductors (MIC), a crucial solution for radiation-hard magnets operating in extreme environments, highlighting the challenges of insulation integrity and indirect cooling. In addition, we reviewed key magnetic measurement techniques—including Hall probes, rotating coils, and moving wires—and their role in determining field integrals, harmonics, and axis alignment. These tools are fundamental for the qualification, alignment, and commissioning of high-precision magnet systems. Finally, we addressed the importance of magnetic coupling effects, especially in assemblies involving permanent magnets and electromagnets, and presented mitigation strategies through multi-magnet simulations and cross-check measurements. In the next lecture, we will turn our attention to superconducting accelerator magnets, exploring their unique capabilities and design constraints.

**Acknowledgement**

The author gratefully acknowledges the support of the Magnet Section at the Paul Scherrer Institute, with special thanks to Rebecca Riccioli, Ciro Calzolaio, Alexander Gabard, Carolin Zoller and Giuseppe Montenero for their contributions to the preparation of this lecture and correcting the manuscript. Sincere appreciation is also extended to Marco Buzio, Davide Tommasini, Thomas Zickler and Luca Bottura from CERN, to my ESRF colleagues Gael Le Bec, Chamseddine Benabderrahmane and Joel Chavanne for their valuable discussions and enduring collaboration spanning more than 20 years.